\documentclass[12pt,thmsa]{article}
\usepackage{amssymb}

\usepackage{sw20lart}

\input{tcilatex}
\input tcilatex
\begin{document}

\title{Optical tests of Bell\'{}s inequalities not resting upon the absurd fair
sampling assumption}
\author{Emilio Santos \\
Departamento de F\'{i}sica, Universidad de Cantabria. \\
Santander. Spain.}
\maketitle

\begin{abstract}
A simple local hidden-variables model is exhibited which reproduces the
results of all performed tests of Bell\'{}s inequalities involving optical
photon pairs. For the old atomic-cascade experiments, like Aspect\'{}s, the
model agrees with quantum mechanics even for ideal set-ups. For more recent
experiments, using parametric down-converted photons, the agreement occurs
only for actual experiments, involving low efficiency detectors. Arguments
are given against the fair sampling assumption, currently combined with the
results of the experiments in order to claim a contradiction with local
realism. New tests are proposed which are able to discriminate between
quantum mechanics and a restricted, but appealing, family of local
hidden-variables models. Such tests require detectors with efficiencies just
above 20\%.
\end{abstract}

\section{\protect\smallskip}

\smallskip

\section{Introduction}

Almost forty years after Bell\'{}s work\cite{Bell}, no experiment has been
performed showing a real (loophole-free) violation of a Bell inequality.
Consequently the important question whether nature obeys local realism is
still open. Furthermore a true refutation of local realism, that is a
disproof of the whole family of local hidden variables (LHV) theories, seems
today more dificult than it was believed twenty years ago. The failure is
concealed behind the assertion that there are loopholes in the performed
tests of Bell\'{}s inequalities, due to (allegedly unimportant)
nonidealities of the measuring devices. This state of opinion is rather
unfortunate because it discourages people from making the necessary effort
to close an open question of fundamental relevance. Indeed it is important
not only for pure science but also for the development of future
technologies, like quantum computation. The truth is that none of the
performed experiments has been able to discriminate between quantum
mechanics and the whole family of LHV theories. Therefore the agreement of
these experiments with quantum mechanics leaves untouched the question of
local realism.

The attempt to prove the impossibility of hidden variables has been a
recurrent endeavor since the early days of quantum mechanics. The reason has
been the attempt to feel confortable in spite of the failure to devise a
picture of the world, consistent with the new theory and fitting in the
tradition of (classical) physics, as supported e. g. by Einstein\cite
{Einstein}. That is, a world view where physical systems have properties
independently of any observation (''the moon is there when nobody looks''),
actions propagate in space-time at a speed not greater than that of light
(without ''spooky actions at a distance'') and probabilities appear due to
ignorance, maybe unavoidable, rather than by an essential indeterminacy
(that is ''God does not play dice''). The mainstream of the scientific
community concealed the failure with the statement that the purpose of
physics is not to understand the world, but just to be able to predict the
results of experiments. This lead to a pragmatic approach to quantum
mechanics, supported mainly by Niels Bohr and known as the \textit{%
Copenhagen interpretation}. But actually it consists of giving up any
interpretation\cite{Peres}. Attempts at a deeper understanding of nature (in
particular, avoiding the \textit{essential indeterminacy)} are commonly
known as \textit{hidden variables theories}. Thus, in order to feel
confortable with the absence of a fully satisfactory interpretation, it was
necessary to prove that hidden variables cannot exist. This lead to a number
of impossibility theorems, the most celebrated proven by John von
Neumann\'{}s in 1932 (the view of von Neumann differs from Bohr\'{}s, but we
shall not discuss this point here). All no-hidden-variables theorems
previous to 1965 were shown to be useless by John Bell\cite{Bell2}. In turn
he stated a new theorem\cite{Bell} which apparently excluded the most
interesting families of hidden variables, those which are local.

Most people are (quite corrrectly) fond of quantum mechanics and interpret
(incorrectly) Bell\'{}s theorem as proving the impossibility of a local
realistic interpretation of nature, even without the need of specific
experiments (e.g. tests of Bell\'{}s inequalities). This is a theoretical
prejudice which support the claim that local realism has been empirically
refuted, in spite of the fact that there are LHV models compatible with all
performed tests of Bell\'{}s inequalities (see below, section 2.)

The failure to get an empirical disproof of local realism, after many years
of attempts, shows that the task is extremely difficult, if not impossible.
This contrasts with the perceived wisdom that quantum mechanics predicts
highly nonlocal effects. (A good illustration of the difficulties involved
is Fry experiment\cite{Fry}. It is the most careful proposal for a
loophole-free test of Bell\'{}s inequalities, it was published in 1995 but
not yet performed eight years later). As time elapses without a true
refutation of local realism, more is reinforced the historical analogy with
the difficulty of making a \textit{perpetuum mobile}, which gave support to
the principles of thermodynamics, or the difficulty of measuring the
absolute velocity of Earth, which lead to relativity theory. That is, time
is reinforcing the conjecture that there are fundamental principles
preventing the violation of a Bell inequality\cite{Santos},\cite{Percival}.

An apparently unsurmontable difficulty for the stated conjecture is that
Bell\'{}s theorem seems to imply that either quantum mechanics is wrong or
local realism is untenable. But the difficulty is not so strong as is
currently assumed. We must take into account that quantum mechanics contains
two quite different ingredients: 1) the formalism (including the equations),
and 2) the postulates of measurement (e. g. ''the possible results of
measuring an observable are the eigenvalues of the associated operator'').
It is a fact that both ingredients are required for the proof of Bell\'{}s
theorem, but only the quantum formalism has been supported (beyond any
reasonable doubt) by the experiments. In contrast most of the postulates of
the measurement are unnecessary, if not absurd, a claim which I hope is
shared by people who today reject any theory of measurement not derived from
the quantum formalism itself.

In summary, it is not obvious that quantum mechanics and local realism are
incompatible, and the success of the former does not refute the latter. As a
consequence new experimental tests of LHV theories are urgently needed.

\section{Local hidden-variables model for optical photon experiments}

Early tests of Bell\'{}s inequalities consisted of measuring the
polarization correlation of entangled photon pairs produced in atomic
cascades\cite{CS}. The first experiment of this type was performed in 1972%
\cite{FC}, and the most celebrated, and sophisticated, was made by Aspect et
al.\cite{Aspect}. Recent tests have used photon pairs produced in non-linear
crystals by the process of parametric down-conversion (PDC)\cite{pdc}. In
most experiments the quantity measured is the polarization correlation of
the photons and we shall consider only these experiments here, although the
argument that follows applies more generally. The set-up consists of a
source of entangled photon pairs, two collecting lens systems, placed on the
oposite sides of the source, plus a polarization analyzer and a detector on
each side. (Experiments using two-channel polarizers and two detectors on
each side will be considered later.) The quantum prediction for single
rates, R$_{1}$ or R$_{2}$, and coincidence rates, R$_{12}$, are\cite{CS},
with some simplifications made for the sake of clarity, 
\begin{eqnarray}
R_{1} &=&R_{2}=\frac{1}{2}\eta R_{0},  \nonumber \\
R_{12}\left( \phi \right) &=&\frac{1}{4}\eta ^{2}R_{0}\left[ 1+V\cos \left(
2\phi \right) \right] ,  \label{1}
\end{eqnarray}
where R$_{0\text{ }}$is the production rate of photon pairs in the source, $%
\eta $ is the overall detection efficiency (which takes into account the
collection efficiency, the losses in the polarizers, lenses, etc, and the
quantum efficiency of the photon detectors) and $\phi $ the angle between
the polarization planes of the analyzers.

In recent PDC\ experiments it is common to use two-channel polarizers with
two detectors inserted behind each polarizer. For every detection in one
beam (conventionally called the\textit{\ signal} beam), a time window is
open for detection in the other (\textit{idler}) beam. The predictions of
quantum mechanics for these experiments may be written 
\begin{equation}
R_{++}\left( \phi \right) =R_{--}\left( \phi \right) =\frac{1}{2}\eta
R_{1}\left[ 1+V\cos \left( 2\phi \right) \right] ,\;R_{+-}\left( \phi
\right) =R_{-+}\left( \phi \right) =R_{++}\left( \phi +\frac{\pi }{2}\right)
.  \label{20}
\end{equation}

For ideal set-up (in particular perfect polarizers and detectors) V = $\eta
=1$ but, due to the nonideality of polarizers, V usually does not surpass
0.97 and, due to nonideality of the idler detector, $\eta $ is less than
0.20 (the efficiency of the detector in the signal beam is irrelevant). Both
quantities, $\eta $ and V, may be measured in independent experiments,
although other nonidealities of the set-up turn down their values when used
in $\left( \ref{20}\right) .$ With reasonable estimates for these effects
the results of the performed experiments confirm quantum theory.

A local hidden variables (LHV) model \cite{Bell} for the experiment consists
of three functions, $\rho (\lambda )$, $P_{1}(\lambda ,\phi _{1})$ and $%
P_{2}(\lambda ,\phi _{2}),$ fulfilling the conditions 
\begin{equation}
\rho (\lambda )\geq 0,\;\;P_{j}(\lambda ,\phi _{j})\geq 0,\;j=1,2,  \label{4}
\end{equation}
\begin{equation}
\int \rho (\lambda )d\lambda =1,\;P_{j}(\lambda ,\phi _{j})\leq 1,
\label{4a}
\end{equation}
and such that 
\begin{eqnarray}
\frac{R_{1}}{R_{0}} &=&\int \rho (\lambda )P_{1}(\lambda ,\phi _{1})d\lambda
,\;\frac{R_{2}}{R_{0}}=\int \rho (\lambda )P_{2}(\lambda ,\phi _{2})d\lambda
,  \nonumber \\
\frac{R_{12}(\phi )}{R_{0}} &=&\int \rho (\lambda )P_{1}(\lambda ,\phi
_{1})P_{2}(\lambda ,\phi _{2})d\lambda ,\;\phi _{1}-\phi _{2}=\phi .
\label{5}
\end{eqnarray}
Here $\lambda $ collectively labels a set of parameters known as ''the
hidden variables''and $\phi _{1}$ ($\phi _{2}$) is the angle of the
polarization plane of the first (second) analyzer with respect to the
horizontal. For later convenience I have separated the homogeneous, $\left( 
\ref{4}\right) ,$ from the inhomogeneous, $\left( \ref{4a}\right) ,$
constraints, the former (the latter) with the property that multiplying
their left hand sides times an arbitrary real number, they remain (may not
remain) true. (For experiments with two channel polarizers we may use two
pairs of functions so that $P_{j-}(\lambda ,\phi _{j})=P_{j+}(\lambda ,\phi
_{j}+\pi /2)).$

In the early papers on the subject it was frequent to assert that LHV models
for the experiments, although possible, would be necessarily contrived\cite
{CH},\cite{CS}. Today we know that this is not the case, as is shown by the
following model. In quantum mechanics, a pair of photons entangled in
polarization may be represented by the state vector 
\begin{equation}
\mid \psi \rangle =\frac{1}{\sqrt{2}}\left( a_{H}^{\dagger }b_{V}^{\dagger
}+a_{V}^{\dagger }b_{H}^{\dagger }\right) \mid vacuum\rangle ,  \label{13}
\end{equation}
where H (V) labels horizontal (vertical) polarization and $a^{\dagger
}(b^{\dagger })$ are creation operators of photons in the first (second)
beam. If we search for a LHV model, it is natural to attach two-dimensional
polarization vectors (in the plane perpendicular to the wave-vectors) to the
incoming light signals, such as 
\begin{equation}
\mathbf{\alpha }\equiv \left( \alpha _{H},\alpha _{V}\right) ,\;\mathbf{%
\beta }\equiv \left( \beta _{V},\beta _{H}\right) .  \label{14}
\end{equation}
The scalar product of these two vectors is the model analog of the
probability amplitude associated to $\left( \ref{13}\right) ,$ whence the
probability should be the square of that scalar product. This suggests a
probability function, $\rho $ (see eq.$\left( \ref{4}\right) )$ of the form 
\begin{equation}
\rho \varpropto (\mathbf{\alpha \cdot \beta )}^{2}\varpropto \cos ^{2}\left(
\lambda _{1}-\lambda _{2}\right) ,  \label{15}
\end{equation}
where the hidden variables $\lambda _{1}$ and $\lambda _{2}$ are angles, say
with the horizontal, defining the directions of the polarization vectors of
the photons. It is easy to see that, after normalization, this gives 
\begin{equation}
\rho =\frac{1}{\pi ^{2}}\left[ 1+\cos \left( 2\lambda _{1}-2\lambda
_{2}\right) \right] .  \label{6}
\end{equation}

Now we shall choose the probabilities, $P_{j}(\lambda ,\phi _{j}),$ of
detection of a photon after passage through the corresponding polarizer. In
a classical theory an electromagnetic signal, polarized in the plane forming
an angle $\lambda _{j}$ with respect to the horizontal, would be divided in
the polarizer according to Malus' law, that is a large intensity would go to
the upper channel if $\lambda _{j}$ is close to the angle, $\phi _{j},$ of
the polarizer (and similarly to the lower channel if $\lambda _{j}$ is close
to $(\phi _{j}+\pi /2))$. Thus it is natural to assume that a detection
event would be more probable if the intensity is large, which suggests that $%
P_{j}(\lambda ,\phi _{j})$ should decrease with increasing value of the
angle $\left| \lambda _{j}-\phi _{j}\right| $. A simple expression which
fits in these conditions is the following one (but the prediction of the
model depends but slightly of the details of that expression, see below ) 
\begin{eqnarray}
P_{j}(\lambda _{j},\phi _{j}) &=&\beta \text{ if }\left| \lambda _{j}-\phi
_{j}\right| \leq \gamma \;\left( \func{mod}\pi \right) \;  \nonumber \\
&&0\text{ otherwise.}  \label{6a}
\end{eqnarray}
By the nature of polarizers the angle $\phi _{j}$ is equivalent to $\phi
_{j}+\pi $, whence the periodicity $\pi .$ For the validity of the model it
is crucial that $\beta \leq 1$ in order that conditions $\left( \ref{4a}%
\right) $ are fulfilled. In this form we obtain a LHV model, not at all
artificial, reproducing the quantum predictions, eqs.$\left( \ref{1}\right)
, $ for all performed experiments involving optical photons.

In fact, inserting eqs.$\left( \ref{6}\right) $ and $\left( \ref{6a}\right) $
in $\left( \ref{5}\right) ,$ it is easy to see that the model leads to 
\begin{equation}
\frac{R_{12}}{R_{1}}=\frac{2\beta \gamma }{\pi }\left[ 1+\frac{\sin
^{2}(2\gamma )}{4\gamma ^{2}}\cos \left( 2\phi \right) \right] ,  \label{7a}
\end{equation}
which reproduces the quantum formula $\left( \ref{1}\right) $ if we assume 
\begin{equation}
V=\frac{\sin ^{2}\left( 2\gamma \right) }{4\gamma ^{2}},\;\eta =\frac{%
4\gamma \beta }{\pi }.  \label{7}
\end{equation}
(It may be realized that using functions P$_{j}$ diferent from $\left( \ref
{6a}\right) ,$ but depending only on$\left| \lambda _{j}-\phi _{j}\right| ,$
would change just the constant factor in front of cos(2$\phi )$ in $\left( 
\ref{7a}\right) ).$ The constraint $\beta \leq 1$ shows that the model may
agree with the quantum predictions only if 
\begin{equation}
V\leq \frac{\sin ^{2}\left( \pi \eta /2\right) }{\left( \pi \eta /2\right)
^{2}}\simeq 1-\frac{\pi ^{2}\eta ^{2}}{12}.  \label{8}
\end{equation}
Typical values in actual experiments are $\eta \lesssim 0.2,$ V $\lesssim
0.97$ (if raw data are used, see next section), so that the LHV model is
compatible with the empirical results. A detailed comparison with a recent
experiment is made in section 5.

\section{The meaning of Bell\'{}s inequalities, homogeneous and inhomogeneous
}

Bell\'{}s theorem has two parts. The first one is the derivation of some
inqualities which hold true for any LHV model. The second part consists of
providing an example of experiment where the quantum-mechanical predictions
violate a Bell inequality. I have no criticism to the first part, but I
claim that the second part is confused. In the current ''proofs'' of
Bell\'{}s theorem, the second part consists of exhibiting a vector and
several projection operators in the Hilbert space of a two-particle system
which, via the standard ''postulates'' of measurement, violate a Bell
inequality. My claim is that a correct proof should not involve the
postulates of measurement. It should consists of the detailed proposal of an
experiment where the predictions of quantum theory violate a Bell
inequality, without any possible loophole. Therefore it seems to me that the
first paper close to being a real proof of the second part of Bell\'{}s
theorem is Fry proposed experiment\cite{Fry}. However, to be sure that an
experiment truly violates a Bell inequality it would be necessary to perform
it. Consequently, in my view Bell\'{}s 1964 work actually consisted of the
derivation of inequalities fulfilled by the whole family of LHV theories,
plus the \textit{suggestion }that these inequalities might be used to tests
local realism (i. e. LHV theories) vs. quantum mechanics. (This is not an
underestimation of Bell\'{}s work, which I believe is one the greatest
achievements in the history of quantum mechanics.)

From eqs.$\left( \ref{4}\right) $ to $\left( \ref{5}\right) $ it is possible
to derive inequalities necessarily fulfilled by \textit{every} LHV model of
the experiments. An example is the (CH) inequality derived by Clauser and
Horne\cite{CH} which I shall write, for a particular choice of polarization
orientations, in the form 
\begin{equation}
\frac{3R_{12}(\phi )-R_{12}(3\phi )}{R_{1}+R_{2}}\leq 1.  \label{CH}
\end{equation}
This inequality is inhomogeneous, in the sense that it compares single rates
with coincidence rates. In its derivation it is necessary to use both, the
homogeneous basic inequalities $\left( \ref{4}\right) $ and the
inhomogeneous ones $\left( \ref{4a}\right) .$ Indeed, it is a general result
that every genuine Bell inequality, valid for the whole family of LHV
theories, must be inhomogeneous\cite{Santos}. However, inhomogeneous
inequalities cannot be tested in optical experiments due to the low
efficiency of optical photon counters. Indeed, the inequality $\left( \ref
{CH}\right) $ cannot be violated in any experiment where the efficiency is
smaller than 82\%. The proposed solution to this problem was to replace one
of the inhomogeneous inequalities $\left( \ref{4a}\right) $ by another one, 
\textit{homogeneous,} which allows deriving testable homogeneous
inequalities. A typical example is the no-enhancement assumption of Clauser
and Horne\cite{CH} 
\begin{equation}
P_{j}(\lambda ,\phi _{j})\leq P_{j}(\lambda ,\infty ),\;j=1,2,  \label{enhan}
\end{equation}
where $P_{j}(\lambda ,\infty )$ is the detection probability with the
polarizer removed. It is possible to prove that this condition, plus $\left( 
\ref{4}\right) $ and the second $\left( \ref{4a}\right) $ leads to an
inequality of the form of $\left( \ref{CH}\right) $, R$_{j\text{ }}$meaning
now the \textit{coincidence} rate with the corresponding polarized removed.
Such inequality is homogeneous and may be violated in experiments with low
efficiency detectors, but the violation does not disprove the whole family
of LHV theories.

Most frequently the substitute for the second $\left( \ref{4a}\right) $ is
not written as an inequality, but expressed as a variant of the ''fair
sampling'' assumption. For instance, ''all photons incident in a detector
have a probability of detection that is independent on whether or not the
photon has passed through a polarizer'', which was used for the derivation
of the inequality tested in the first atomic-cascade experiment\cite{FC}.
Or, ''if a pair of photons emerges from the polarizers the probability of
their joint detection is independent of the angles $\phi _{1}$ and $\phi
_{2} $'', used in the first experimental proposal\cite{CHSH}. Hence the
authors derived the most celebrated \textit{homogeneous }inequality, known
as CHSH, valid for experiments with two-channel polarizers 
\begin{equation}
S=\left| 3E\left( \pi /8\right) -E\left( 3\pi /8\right) \right| \leq 2,\text{
}E\left( \phi \right) =\frac{R_{++}\left( \phi \right) +R_{--}\left( \phi
\right) -R_{+-}\left( \phi \right) -R_{-+}\left( \phi \right) }{R_{++}\left(
\phi \right) +R_{--}\left( \phi \right) +R_{+-}\left( \phi \right)
+R_{-+}\left( \phi \right) },  \label{31}
\end{equation}
where R$_{++}$, R$_{--}$, R$_{+-}$, R$_{-+},$ are coincidence rates in the
four pairs of detectors, taking one detector on each side.

If the quantum prediction $\left( \ref{20}\right) $ is inserted in the CHSH
inequality $\left( \ref{30}\right) $ we obtain 
\begin{equation}
V\leq \sqrt{2}/2\simeq 0.7071,  \label{51}
\end{equation}
which is the inequality actually tested, and violated, in the performed
experiments. In sharp contrast, if we insert the quantum prediction in the
genuine, inhomogeneous, Bell inequality $\left( \ref{CH}\right) $ (taking,
e.g. R$_{++}$ for R$_{12}$) we get

\begin{equation}
\eta \left( 1+\sqrt{2}V\right) \leq 2,  \label{50}
\end{equation}
which could be violated only if both, $\eta $ and V, were close enough to
unity.

The sharp contrast between the tested inequality $\left( \ref{51}\right) $
and the Bell inequality $\left( \ref{50}\right) $ urges me to stress that
Bell\'{}s inequalities refer to \textit{correlations }(e. g. in
polarization) between \textit{distant} measurements (in the most strong,
relativistic, form space-like should be substituted for distant). In a test
of Bell\'{}s inequalities, the correlation is quantified by the parameter V,
but the distance is related to the position measurement, which involves $%
\eta .$ Consequently I claim that replacing the test of the inequality $%
\left( \ref{50}\right) $ by the test of $\left( \ref{51}\right) $ amounts to
losing an essential condition of the test.

The parameter V, related to the polarization, is a wave property, whilst the
detection efficiency $\eta $, related to localization, is a particle
property. Thus we might interpret the inequality $\left( \ref{50}\right) $
as forbiding that photons behave as (localized) particles and (delocalized)
waves at the same time, whilst $\left( \ref{51}\right) $ just tests the wave
property. Indeed, in optical photons the wave properties manifest most
strongly and it is the ''particle'' parameter $\eta $ which is usually
small. In contrast, it is known that in gamma rays, which behave more
strongly as particles, it is the polarization which is difficult to measure
accurately, whilst 100\% efficient detection is achievable. Thus, the high
values of V (close to unity) found in optical experiments tell us, simply,
that optical photons are extended objects (waves) rather than localized
objects (particles). A more quantitative statement should replace extended
(localized) by bigger (smaller) than atoms.

\section{Criticism of the fair sampling assumption}

In order to see whether the ''fair sampling'' assumption is plausible, as is
usually assumed, it is illustrative to return to the old atomic-cascade
experiments, like Aspect\'{}s one$\cite{Aspect}$. In these experiments the
angular correlation of the photons emitted by the source is poor due to the
three-body character of the atomic decay. Thus the quantum prediction eq.$%
\left( \ref{1}\right) $ may be written in the form 
\begin{eqnarray}
R_{1} &=&R_{2}=\frac{1}{2}\frac{\Omega }{4\pi }\eta ^{\prime }R_{0}, 
\nonumber \\
R_{12}\left( \phi \right) &=&\frac{1}{4}\eta ^{\prime 2}\left( \frac{\Omega 
}{4\pi }\right) ^{2}R_{0}\left[ 1+V^{\prime }F\left( \alpha \right) \cos
\left( 2\phi \right) \right] ,  \label{1a}
\end{eqnarray}
where the new parameters $\eta ^{\prime }$ and V' are related to
nonidealities of detectors and polarizers, respectively, $\alpha $ is the
half angle of the apertures (assumed equal) and $\Omega $ the solid angle
covered by them, that is 
\begin{equation}
\Omega =2\pi \left( 1-\cos \alpha \right) .  \label{2a}
\end{equation}
(For the sake of clarity we have ignored a factor close to unity in front of
the right hand side of the second eq.$\left( \ref{1a}\right) ,$ but this
does not affects the argument that follows). The function $F\left( \alpha
\right) $ takes into account that the polarization correlation of a photon
pair decreases when the angle between their wave-vectors departs from $\pi $%
. In the most frequent case of a 0-1-0 cascade (that is, both the initial
and final states of the atom having zero spin) the form of the function is 
\begin{equation}
F\left( \alpha \right) =1-\frac{2}{3}\left( 1-\cos \alpha \right) ^{2}.
\label{3a}
\end{equation}
It is easy to see that the LHV model of section 2 is compatible with the
quantum predictions $\left( \ref{1a}\right) $ even for ideal set-up, that is 
$\eta ^{\prime }=V^{\prime }=1.$

The point that I want to stress is that in atomic-cascade experiments the
''fair sampling assumption'', that the ensemble of detected photon pairs is
representative of the whole ensemble of pairs emitted from the source,
contradicts quantum mechanics. In fact, quantum theory predicts that the
polarization correlation of a photon pair decreases when the angle between
their wave-vectors departs from $\pi .$ As a consequence the correlation is
higher for the photons actually detected than for the average pairs produced
in the source, a fact giving rise to the factor F($\alpha )$ in eq.$\left( 
\ref{1}\right) $. It may be said that fair sampling refers to the ensemble
of pairs such that both enter the apertures, and this might be applied to
both atomic-cascade and PDC experiments. But such ensemble is nonsense
according to quantum mechanics, in the same way as the ''ensemble of photons
passing the upper slit'' is nonsense in a two-slit experiment\cite
{Santosreply}. In summary, fair sampling is either false or nonsense in
quantum mechanics.

In the context of LHV theories the fair sampling assumption is, simply,
absurd. In fact, the starting point of \textit{any} hidden variables theory
is the hypothesis that quantum mechanics is not complete, which essentially
means that states which are considered identical in quantum theory may not
be really identical. For instance if two atoms, whose excited states are
represented by the same wave-function, decay at different times, in quantum
mechanics this fact may be attributed to an ''essential indeterminacy'',
meaning that identical causes (identical atoms) may produce different
effects (different decay times). In contrast, the aim of introducing hidden
variables would be to explain the different effects as due to the atomic
states not being really identical, only our information (encapsuled in the
wave-function) being the same for both atoms. That is, \textit{the essential
purpose} of hidden variables is to attribute differences to states which
quantum mechanics may consider identical. Therefore it is absurd to use the
fair sampling assumption -which rests upon the identity of all photon pairs-
in the test of LHV theories, because that assumption excludes hidden
variables \textit{a priori}.

For similar arguments it is not allowed to subtract accidental coincidences,
but the raw data of the experiments should be used. In fact, what is
considered accidental in the quantum interpretation of an experiment might
be essential in a hidden variables theory. Actually any interpretation of
quantum theory in terms of hidden variables should take into account the
existence of a fundamental noise (quantum fluctuations). The quantum
formalism has been developed containing (very clever) rules which allow an
efficient removal of that noise in the calculations, e. g. the use of normal
ordering in quantum field theory. But a hidden variables theory might relate
noise in measurements (e. g. dark rate in detection) to quantum
fluctuations, so attributing a fundamental character to a part, at least, of
the experimental noise. This is the reason why uncorrected (raw) data should
be used in the interpretation of experiments testing LHV theories.

\section{New test of local hidden variables with optical photon experiments}

In experiments with optical photons the main loophole for a disproof of
local realism derives from the low efficiency (and/or high dark rate) of
available photon counters. It is well known that efficiencies higher than
about 80\% (with no noise) are required for a loophole-free experiment.
Apparently such photon counters will not be available in the near future
and, consequently, it seems appropriate to perform new experiments which
might provide some indication about the results to be expected when the
efficiency of detectors increases. As we have remarked, making new
experiments, whose interpretation requires the fair sampling assumption,
adds very little to our knowledge. The alternative is to test quantum
mechanics against restricted families of LHV theories and I propose models
of the type presented in the previous section.

Actually, the proposal of testing restricted families of LHV theories is not
new, because introducing assumptions, e. g. $\left( \ref{enhan}\right) $, in
addition to eqs.$\left( \ref{4}\right) $ and $\left( \ref{4a}\right) $ in
order to get testable inequalities, amounts to restricting the family of LHV
theories to be tested. Such additional assumptions, allegedly plausible,
have been used for the interpretation of all performed tests, as was
commented in section 3. However, as plausibility is not a scientific
criterion, it seems to me that a proposal where it is clearly exhibited the
restricted family of LHV theories to be tested, is less misleading than
using the word ''plausible''. Such a proposal is what is made in the
following.

We consider a family of LHV models with two angular hidden variables $%
\left\{ \lambda _{1},\lambda _{2}\right\} $, a probability density, $\rho ,$
of the photon pairs created in the source depending only on $\left| \lambda
_{1}-\lambda _{2}\right| $ and probability functions, $P_{j}$, depending on $%
\left| \lambda _{j}-\phi _{j}\right| $. It would be desirable to define a
criterion of closeness with quantum predictions and, after that, to optimize
the functions $\rho $ and P$_{j}$ in order to get the best LHV model of the
said family, but this program will be left for the future. Here we shall
consider two particular models appropriate for low an high efficiencies,
respectively. More specifically for situations where the inequality $\left( 
\ref{8}\right) $ is either slightly of strongly violated. In both case we
shall use functions $P_{j}$ of the form $\left( \ref{6a}\right) .$

For low efficiency we argue as follows. The probability density $\rho $ may
be written as a Fourier series 
\begin{equation}
\rho =\frac{1}{\pi ^{2}}\left[ 1+\sum_{n=1}a_{n}\cos \left( 2n\lambda
_{1}-2n\lambda _{2}\right) \right] ,  \label{20a}
\end{equation}
normalized in the interval $\left[ 0,\pi \right] .$ The simple model of
section 2 corresponds to a$_{1}$ = 1, a$_{n}$ = 0 for n \TEXTsymbol{>} 1.
This model agrees with quantum predictions only if the inequality $\left( 
\ref{8}\right) $ holds true. If the inequality is violated, we need a
function $\left( \ref{20a}\right) $ with a$_{1}$ \TEXTsymbol{>} 1 which, by
the positivity of $\rho ,$ necessarily requires that some a$_{n}$ with n 
\TEXTsymbol{>} 1 are not zero. The most simple choice is 
\begin{equation}
\rho =\frac{1}{\pi ^{2}}\left[ 1+\left( 1+\varepsilon \right) \cos \left(
2\lambda _{1}-2\lambda _{2}\right) +\varepsilon \cos \left( 4\lambda
_{1}-4\lambda _{2}\right) \right] .  \label{21}
\end{equation}
which involves an adjustable parameter, $\varepsilon \in \left[ 0,\frac{1}{3}%
\right] $ (compare with eq.$\left( \ref{6}\right) ).$ The proposed density
is non-negative definite, being equivalent to 
\begin{equation}
\rho =\frac{2}{\pi ^{2}}\left[ \left( 1-3\varepsilon \right) \cos ^{2}\left(
\lambda _{1}-\lambda _{2}\right) +4\varepsilon \cos ^{4}\left( \lambda
_{1}-\lambda _{2}\right) \right] .  \label{22}
\end{equation}
Putting eqs.$\left( \ref{6a}\right) $ and $\left( \ref{21}\right) $ into eqs.%
$\left( \ref{5}\right) $ we get 
\begin{equation}
\frac{R_{12}}{R_{1}}=\frac{2\beta \gamma }{\pi }\left[ 1+\left(
1+\varepsilon \right) \frac{\sin ^{2}(2\gamma )}{4\gamma ^{2}}\cos \left(
2\phi \right) +\varepsilon \frac{\sin ^{2}(4\gamma )}{16\gamma ^{2}}\cos
\left( 4\phi \right) \right] .  \label{23}
\end{equation}
This model prediction generalizes $\left( \ref{7a}\right) $ (see section 3)
and agrees with it if $\varepsilon =0$.

Tests of the LHV model against quantum mechanics require new experiments
with higher values of the parameters $\eta $ and/or V, e. g. using
parametric down converted photons. This means not too small quantum
efficiencies of the photon counters, but also high collecting apertures for
the idler photon in each pair. In contrast the apertures for the signal
photon may be arbitrarily small, with the limitation of not reducing too
much the single counting rate in this beam. For such experiments our LHV
model predicts deviations of the function R$_{12}\left( \phi \right) $ from
the quantum prediction $\left( \ref{1}\right) $, because the parameter $%
\varepsilon $ cannot be zero (see $.\left( \ref{23}\right) )$. In fact, the
condition $\beta \leq 1$ gives in this case (using the second eq.$\left( \ref
{7}\right) )$%
\begin{equation}
V\leq \left( 1+\varepsilon \right) \frac{\sin ^{2}(\pi \eta /2)}{\pi
^{2}\eta ^{2}/4},  \label{26}
\end{equation}
so that $\varepsilon >0$ whenever the inequality $\left( \ref{8}\right) $ is
violated. Thus the LHV model predicts deviations from the quantum formula $%
\left( \ref{1}\right) ,$ which may be tested experimentally, consisting in
the presence of a cos($4\phi )$ term in the fit of the empirical curve R$%
_{12}$($\phi ).$

We define the deviation between the quantum and model predictions by 
\begin{equation}
\delta \equiv \left\langle \left[ \frac{R_{12}^{Q}(\phi )}{\left\langle
R_{12}^{Q}(\phi )\right\rangle }-\frac{R_{12}^{LHV}(\phi )}{\left\langle
R_{12}^{LHV}(\phi )\right\rangle }\right] ^{2}\right\rangle ^{1/2},
\label{26a}
\end{equation}
where \TEXTsymbol{<}\TEXTsymbol{>} means average over angles. Taking $\left( 
\ref{23}\right) $ and $\left( \ref{26}\right) $ into account, we get 
\begin{equation}
\delta \geq \frac{\sqrt{2}\varepsilon }{2}\frac{\sin ^{2}\left( \pi \eta
\right) }{\left( \pi \eta \right) ^{2}}\geq \frac{\sqrt{2}}{2}\cos
^{2}\left( \pi \eta /2\right) \left[ V-\frac{\sin ^{2}(\pi \eta /2)}{(\pi
\eta /2)^{2}}\right] .  \label{26b}
\end{equation}
This is our proposed Bell inequality. In the empirical test, the parameter V
may be obtained from a fit of the empirical data to the quantum prediction
eq.$\left( \ref{1}\right) .$ The inequality requires only the measurements
of coincidence rates and detection efficiency.

A related inequality may be obtained as follows. We define two parameters V$%
_{A}$ and V$_{B}$ : 
\begin{equation}
V_{A}=\frac{R_{12}(0)-R_{12}(\pi /2)}{R_{12}(0)+R_{12}(\pi /2)},V_{B}=\sqrt{2%
}\frac{R_{12}(\pi /8)-R_{12}(3\pi /8)}{R_{12}(\pi /8)+R_{12}(3\pi /8)},
\label{27}
\end{equation}
which according to the quantum prediction are both equal to the parameter V
of eq.$\left( \ref{20}\right) .$ In contrast, in our model (with $%
\varepsilon \neq 0)$ we get from $\left( \ref{23}\right) $ 
\begin{equation}
V_{B}=V=\left( 1+\varepsilon \right) \frac{\sin ^{2}(\pi \eta /2\beta )}{%
(\pi \eta /2\beta )^{2}},\;V_{A}=V_{B}\left[ 1+\varepsilon \frac{\sin
^{2}(\pi \eta /\beta )}{(\pi \eta /\beta )^{2}}\right] ^{-1}.  \label{28}
\end{equation}
Taking the constraint $\beta \leq 1$ into account, this implies 
\begin{equation}
\frac{V_{B}}{V_{A}}\geq 1+\cos ^{2}\left( \pi \eta /2\right) \left[ V_{B}-%
\frac{\sin ^{2}(\pi \eta /2)}{(\pi \eta /2)^{2}}\right] ,  \label{29}
\end{equation}
which may be tested by measuring the three quantities $V_{A},V_{B}$ (or $V$)
and $\eta .$ Quantum predictions should violate the inequality for high
enough $V$ and $\eta .$ This inequality has the advantage that the
parameters V$_{A}$ and V$_{B}$ have been measured in most experiments, which
allows tests using the results already reported (see below). However, this
inequality is only useful for relatively small detection efficiencies, where
the model resting upon eq.$\left( \ref{21}\right) $ is appropriate. But the
inequality does not hold true for the model described below (see eq.$\left( 
\ref{35}\right) $), appropriate for high efficiencies.

Our model may be extended to experiments with two-channel polarizers using
two pairs of functions such that $P_{j-}(\lambda ,\phi _{j})=P_{j+}(\lambda
,\phi _{j}+\pi /2)$, which leads to the prediction 
\begin{equation}
R_{+-}\left( \pi /2+\phi \right) =R_{-+}\left( \pi /2+\phi \right)
=R_{++}\left( \phi \right) =R_{--}\left( \phi \right) ,  \label{30}
\end{equation}
the latter given by eq.$\left( \ref{23}\right) .$ It is not difficult to see
that, with this choice, V$_{A\text{ }}$ coicides with the visibility of any
of the functions $\left( \ref{30}\right) $ (if they are equal, or their
average if they are not) and V$_{B\text{ }}$ coincides with $\sqrt{2}/4$
times the quantity S defined by eq.$\left( \ref{31}\right) ,$ usually
measured in the tests of Bell\'{}s inequality. This allows testing $\left( 
\ref{29}\right) $ in experiments with two-channel polarizers and some
performed experiments are quite close to achieving a test of the inequality.
For instance the experiment of Kurtsiefer et al.\cite{Kurt} reports a value
S = 2.6979 $\pm $ $0.0034$, leading to V$_{B}$ = 0.9539 $\pm $ $0.00012,$
with $\eta =$ 0.214. Thus the inequality $\left( \ref{50}\right) $ (with V$%
_{B\text{ }}$ substituted for V) is fulfilled, which means that the
experiment is unable to discriminate between quantum mechanics and our
model. However, an increase of S by only 1\% could provide a valid test of
eq.$\left( \ref{29}\right) .$

Another possibility, in experiments with two-channel polarizers, would be to
see whether the sum 
\begin{equation}
R_{tot}(\phi )=R_{++}\left( \phi \right) +R_{--}\left( \phi \right)
+R_{+-}\left( \phi \right) +R_{-+}\left( \phi \right)  \label{34}
\end{equation}
is independent of $\phi $. In our model, using eq.$\left( \ref{30}\right) ,$
the quantity R$_{tot}$ is proportional to 1 + $\varepsilon \cos \left( 4\phi
\right) $, so that deviations from rotational invariance are expected of
order $\varepsilon $ ($\varepsilon $ may be related to the measurable
parameters $\eta ,V_{A\text{ }}$ and V$_{B}$ (or$V$) by the second eq.$%
\left( \ref{28}\right) )$ . Measurements of the deviation from rotational
invariance of the quantity R$_{tot}$ have been recently proposed as a test
of the fair sampling assumption\cite{Adenier}.

It might be believed that models of the type presented in this section would
depart strongly from the quantum predictions when the detection efficiency
increases, for instance above 80\%, but this is not the case. For high
values of $\eta $ and/or V a good LHV model, of the same type, consists of
using the probability function 
\begin{equation}
\rho =\frac{1}{\sqrt{2\pi ^{3}}\sigma }\exp \left( \frac{\lambda ^{2}}{%
2\sigma ^{2}}\right) \left( \func{mod}\pi \right) ,\lambda =\lambda
_{1}-\lambda _{2}.  \label{35}
\end{equation}
This is a periodic function which may be expanded in Fourier series to give 
\begin{equation}
\rho =\frac{1}{\pi ^{2}}\left[ 1+2\sum_{n=1}\exp \left( -2n^{2}\sigma
^{2}\right) \cos \left( 2n\lambda \right) \right] ,  \label{36}
\end{equation}
provided that $\sigma $ is small enough so that the gaussian function in eq.$%
\left( \ref{35}\right) $ is negligible for $\lambda =\pi /2.$ Using again $%
\left( \ref{6}\right) $ for the detection probabilities we get a prediction
(for $\beta =1)$%
\begin{equation}
\frac{R_{12}}{R_{1}}=\frac{2\gamma }{\pi }\left[ 1+2\sum_{n=1}\exp \left(
-2n^{2}\sigma ^{2}\right) \frac{\sin ^{2}\left( 2n\gamma \right) }{\left(
2n\gamma \right) ^{2}}\cos \left( 2n\phi \right) \right]  \label{37}
\end{equation}
This is very close to the quantum prediction even for quite high detection
efficiency, in particular using $\sigma =\pi /18,$ as recently proposed\cite
{Adenier}. For this choice it is possible to get V = 1 with efficiencies up
to $\eta =0.848$ and still the departures from the quantum prediction are
only of a few percent. In fact we get 
\[
V_{A}=0.980,\;V_{B}=0.957,\;\delta =0.047. 
\]
Even closer is the prediction for the function $E\left( \phi \right) $, see
eq.$\left( \ref{31}\right) ,$ where the discrepancy corresponds to a term in
cos$\left( 6\phi \right) $ with a coefficient of order 0.02.

For higher efficiencies the model cannot reproduce the (ideal) quantum
prediction V = 1. In particular, for 100\% efficiency, i. e. $\eta
=1\Rightarrow \gamma =\pi /4,$ we get V $\leq 8/\pi ^{2},$ whatever is the
value of $\sigma .$ But, even in this limit, the model does not depart too
much from quantum mechanics. For instance, with the same choice $\sigma =\pi
/18,$ we get 
\[
V=0.7627,V_{A}=0.8225,\;V_{B}=0.7043, 
\]
in comparison with the ideal quantum prediction V = $V_{A}=V_{B}=1.$ It may
be realized that the value given by that model for the left side of the CH
inequality $\left( \ref{CH}\right) $ is $0.996$, very close to the Bell
limit, 1, although not so close to the quantum prediction with ideal
polarizers, 1.207. It is remarkable that the function R$_{tot}$($\phi )$, eq.%
$\left( \ref{34}\right) ,$ defined with the prescription eq.$\left( \ref{30}%
\right) ,$ is unity for any angle $\phi $, in exact agreement with the
quantum prediction. It is also interesting that the model gives $%
V_{A}>V_{B}, $ in contrast with what happens with the model of the previous
section, leading to the inequality $\left( \ref{29}\right) .$ Also our
proposed Bell inequality $\left( \ref{26b}\right) $, which was derived for
low detection efficiencies, is here useless its right hand side being zero.
This shows that this inequality is only appropriate for relatively low
detection efficiencies.

In summary, the proposed family of models agrees exactly with quantum
predictions for low enough detection efficiency, and departs when the
efficiency increases. The departure manifests in that the curve $%
R_{12}\left( \phi \right) $ , when Fourier analyzed, either contains terms
in cos(2n$\phi )$ with n \TEXTsymbol{>} 1 or has a lower coefficient in the
cos(2$\phi ),$ or both. The deviations slowly increase with the detection
efficiency up to a few percent for 85\% efficiency. Then more rapidly up to
about 20\% for 100\% efficiency. But neither the inequality $\left( \ref{29}%
\right) $ is necessarily violated by the models nor R$_{tot}$($\phi )$, eq.$%
\left( \ref{34}\right) ,$ necessarily exhibits lack of rotational symmetry.
In any case, for not too large detection efficiency (say below 30\%) the
inequality $\left( \ref{26b}\right) $ provides a good test of an appealing
family of local hidden variables models vs. quantum mechanics.

\section{Discussion}

Current optical tests of Bell\'{}s inequalities - consisting of measuring a
coincidence counting rate as a function of some angular parameter and
confirming that the curve is a cosinus one with visibility,\ V, above 0.71 -
are useless for the purpose of refuting local realism. In this type of test
it is therefore worthless to improve the experiment by increasing the
distance between the detectors or increasing the statistics in order to
provide violations with more standard deviations.

I propose experiments, easily feasible, which would test a particular, but
interesting, family of local hidden variables theories. The test may be
performed using detectors with not too high efficiency, just above 20\%. On
the other hand the test requires only the measurement of coincidence rates
and it is, therefore, inmune to many sources of noise. It is true that the
refutation of that family would not rule out local realism, because other
kinds of models exist in perfect agreement with quantum predictions up to
about 64\% efficiency\cite{Caser}, but those models are more artificial.

In the proposed experiments there exists the possibility that the standard
quantum predictions are violated, which I claim would not contradict the
hard core of quantum mechanics, that is the formalism. It would disprove
only the standard theory of measurement. It is generally believed that
departures from ideal behaviour of polarizers, detectors, etc. are just
practical problems, which should be solved with the progress of the
technology. But there may be nonidealities of a fundamental character. For
instance, a detailed (quantum) theory of photon counters might show, at high
efficiencies, a nonlinear relation between the detection rate and the
intensity of the incoming light beam, or an unavoidable increase of the dark
rate. The models studied in the present paper show that relatively small
nonidealities might be sufficient to make compatible with LHV theories the
experiments involving entangled photon pairs.

\end{document}